\documentclass[aps,prb,twocolumn,showpacs]{revtex4-1}

\def\vR{{\bf R}}
\def\vk{{\bf k}}
\def\vq{{\bf q}}
\def\vr{{\bf r}}

\def\flip{\uparrow\downarrow}

\usepackage{graphicx,multirow}
\usepackage{CJK}

\begin{document}

\begin{CJK*}{UTF8}{bsmi}
\title{Electron spin-phonon interaction symmetries and tunable spin relaxation in silicon and germanium}
\author{Jian-Ming Tang
(湯健銘)
}
\author{Brian T. Collins}
\affiliation{Department of Physics, University of New Hampshire, Durham, NH 03824-3520, USA}
\author{Michael E. Flatt\'e}
\affiliation{Optical Science and Technology Center and Department of Physics and Astronomy, University of Iowa, Iowa City, IA 52242-1479, USA}

%\date{\today}

\begin{abstract}
Compared with direct-gap semiconductors, the valley degeneracy of silicon and germanium opens up new channels for spin relaxation that counteract the spin degeneracy of the inversion-symmetric system. Here the symmetries of the electron-phonon interaction for silicon and germanium are identified and the resulting spin lifetimes are calculated. Room-temperature spin lifetimes of electrons in silicon are found to be comparable to those in gallium arsenide, however, the spin lifetimes in silicon or germanium can be tuned by reducing the valley degeneracy through strain or quantum confinement. The tunable range is limited to slightly over an order of magnitude by intravalley processes.
\end{abstract}


\maketitle 
\end{CJK*}

\section{Introduction}

The favorable material properties of silicon have permitted it to dominate the microelectronics industry for over half a century, however a new genre of spintronic semiconductor devices,\cite{Wolf2001,Awschalom2002,Awschalom2007} in which spins of electronic carriers are manipulated instead of a charge current, requires long spin transport lengths and coherence times. Although spin injection into nonmagnetic semiconductors was demonstrated over a decade ago,\cite{fiederling_injection_1999,ohno_electrical_1999,hanbicki_efficient_2002,strand_dynamic_2003,adelmann_spin_2005} the recent success at injecting spin-polarized current into silicon\cite{appelbaum_electronic_2007, jonker_electrical_2007, dash_electrical_2009,grenet_spin_2009} suggests incorporation of semiconductor spintronic device concepts into hybrid silicon device architectures.  Polarized spins relax in semiconductors because the spin-orbit interaction entangles orbital and spin degrees of freedom, and thus ordinary scattering from defects or lattice vibrations leads to a loss of spin coherence. In materials without inversion asymmetry the entanglement of spin and orbit manifests as an effective momentum-dependent (internal) magnetic field, causing spin precession and  D'yakonov-Perel' spin relaxation. Within inversion-symmetric materials, such as silicon, the internal magnetic field vanishes, but scattering between states with spin-orbit entangled wave functions leads to Elliott-Yafet spin relaxation. The spin coherence times in silicon are long at low temperature, and the spin-orbit interaction and lattice symmetry reduces spin relaxation rates relative to optically accessible (direct-gap) semiconductors.\cite{kikkawa_resonant_1998,boggess_room-temperature_2000,beschoten_spin_2001} The silicon band structure, however, has multiple valleys that permits low-energy scattering of electrons by large momenta, which allows the Elliott-Yafet process to be more effective.\cite{optical_orientation_1984}  Numerical calculations that include these effects have been successful at explaining the spin lifetime in silicon as a function of temperature.\cite{cheng_theory_2010,li_spin_2011} Tuning the spin lifetime in inversion-asymmetric semiconductors with a single direct gap have largely focused on the electric-field-induced Rashba spin-orbit interaction that shortens the spin lifetime;\cite{lau_tunability_2002,averkiev_spin-relaxation_2006} unaddressed is the potential for new methods of tuning the spin lifetime associated with the valley degeneracy of the semiconductor.

Here we trace that origin of the intrinsic spin lifetime in silicon to the spin flips associated with large momentum transfer events, and disentangle the intervalley contribution to spin relaxation from the intravalley contribution. As the different processes involve different momentum transfers, and for the intrinsic spin-relaxation rate the source of that momentum transfer will be electron-phonon scattering, the various contributions will be associated with specific regions of the phonon dispersion curves. Due to the high symmetry of the crystal lattice, several processes that might have been expected to contribute will be forbidden by symmetry. We provide a full symmetry analysis of the various contributions to the spin-relaxation rate from phonon-mediated scattering. The separation of spin-relaxation mechanisms by momentum transfer also permits a direct calculation of the tuning of spin lifetime possible by splitting the energies of the electron valleys and thus suppressing some of the intervalley scattering effects.  Reducing the valley degeneracy in silicon, through applied strain or the growth of pseudomorphic SiGe quantum wells, will reduce the effect of the spin-orbit interaction on electron scattering, lengthening the spin coherence time and spin transport length. We find an effective tuning range of approximately one order of magnitude.

\section{Electron-Phonon Scattering}
 
The intrinsic spin-relaxation time is determined by electron-phonon scattering. For one-phonon absorption ($+$) and emission ($-$) processes, the scattering probability from $\vk$ to $\vk'$ is
\begin{eqnarray}
M_{\sigma'\sigma}^\pm(\vk',\vk) & = & \left|\left\langle\psi_{\vk'\sigma'},n_\vq\mp 1\left|\hat H^{\rm ep}_\pm\right|\psi_{\vk\sigma},n_\vq\right\rangle\right|^2 ,
\end{eqnarray}
where $\hat H_\pm^{\rm ep}$ is the time-independent part of the electron-phonon interaction Hamiltonian corresponding to absorption or emission, $\vq=\vk'-\vk$, $\sigma$ labels the spin state, and $n_\vq$ is the phonon occupation number.
To evaluate $M_{\sigma'\sigma}^\pm(\vk',\vk)$ for various types of scattering processes in the material,
we use a first-nearest-neighbor $sp^3$ tight-binding model (TBM) with on-site spin-orbit interactions\cite{chadi_spin-orbit_1977} to obtain the wave functions,
\begin{eqnarray}
\psi_{\vk\sigma}(\vr) & = & \frac{1}{\sqrt{N}}\sum_{j,a,l,s} c_{als} e^{i\vk\cdot\vR_{ja}}\phi_{al}(\vr-\vR_{ja})\chi_s ,
\end{eqnarray}
where $N$ is the number of unit cells, $j$ labels unit cells, $a$ labels the two basis atoms
within a unit cell, $l$ labels the atomic orbital bases, $\chi$ is a two-component spinor, $s$ is the spin index, and $\vR_{ja}$ is the position vector of atoms. We choose the spin quantization axis to be aligned with the $z$ axis and determine the coefficients $c_{als}$ by maximizing the expectation value of the spin operator $\langle\hat S_z\rangle$.

%the interaction Hamiltonian is
%\begin{eqnarray}
%\hat H^{\rm ep}_\pm & = & \sum_{j,a}\frac{\partial\hat H}{\partial\vR_{ja}}\cdot\sqrt{\frac{\hbar}{2\rho V\omega}}\sum_\vq\vu_a\hat{b}_{\vq}^\mp e^{\pm i\vq\cdot\vR_{ja}} ,
%\end{eqnarray}
%$\vu_a$ is the unit polarization vector
%$\hat b_\vq^\pm$ are the phonon creation/annihilation operators.

%the scattering rates are
%\begin{eqnarray}
%\Gamma_{\sigma'\sigma}^\pm & = & \frac{2\pi V}{\hbar}\int\frac{d^3\vk'}{(2\pi)^3}M_{\sigma'\sigma}^\pm(\vk',\vk)\delta(E_{\vk'}-E_\vk\mp\hbar\omega) ,
%\label{eq:scattering_rate}
%\end{eqnarray}

\begin{figure}
\includegraphics[width=\columnwidth]{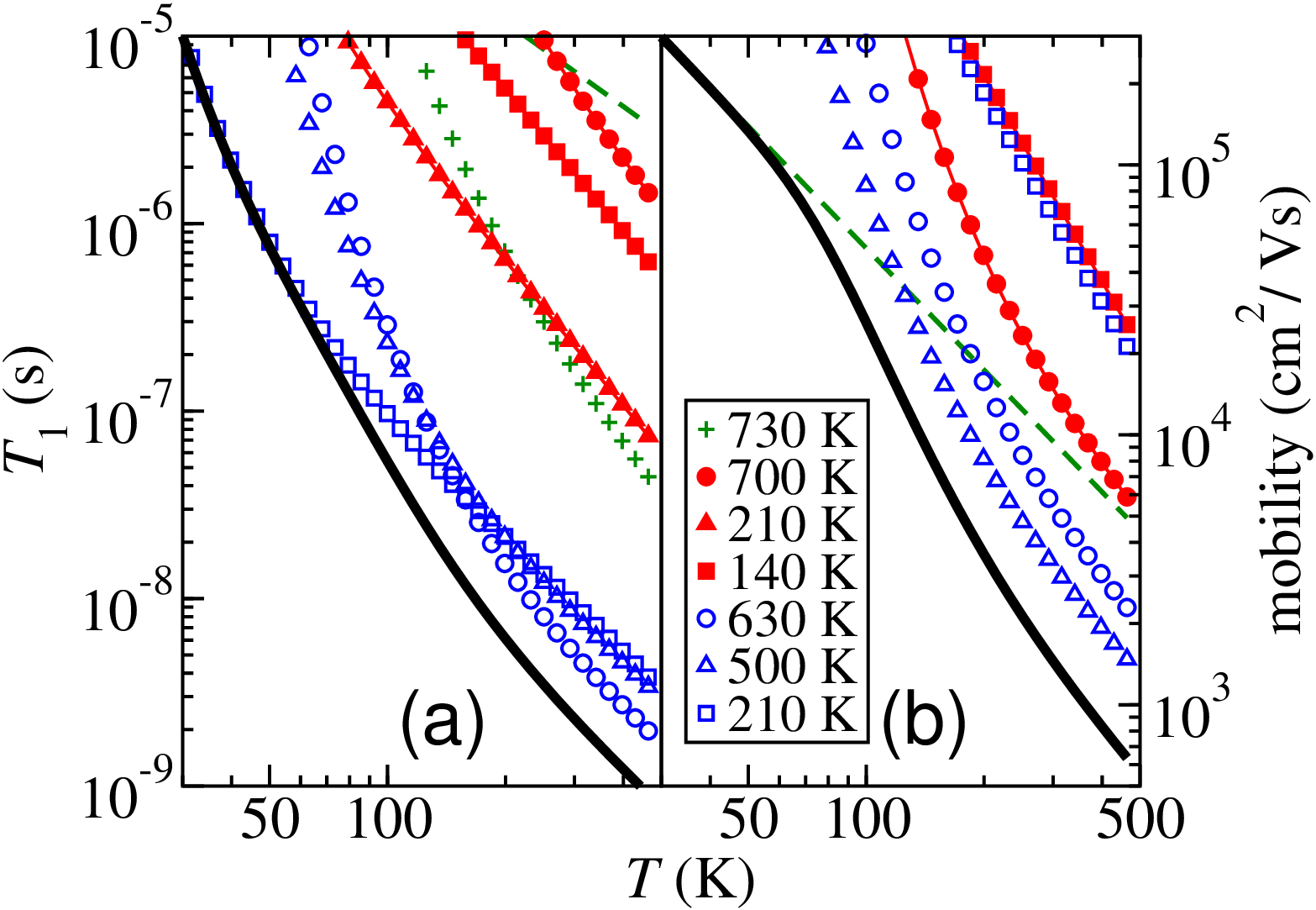}
\caption{(Color online) Electron spin-relaxation time (a) and mobility (b) in bulk silicon shown by black solid lines. Individual scattering processes are intravalley acoustic process (dashed line), intravalley optical ($\Gamma_{25}^+$) process (plus symbol), intervalley $\Delta$ processes (closed symbols), and intervalley $\Sigma$ processes (open symbols).}
\label{fig:si_spin}
\end{figure}

\begin{figure}
\includegraphics[width=\columnwidth]{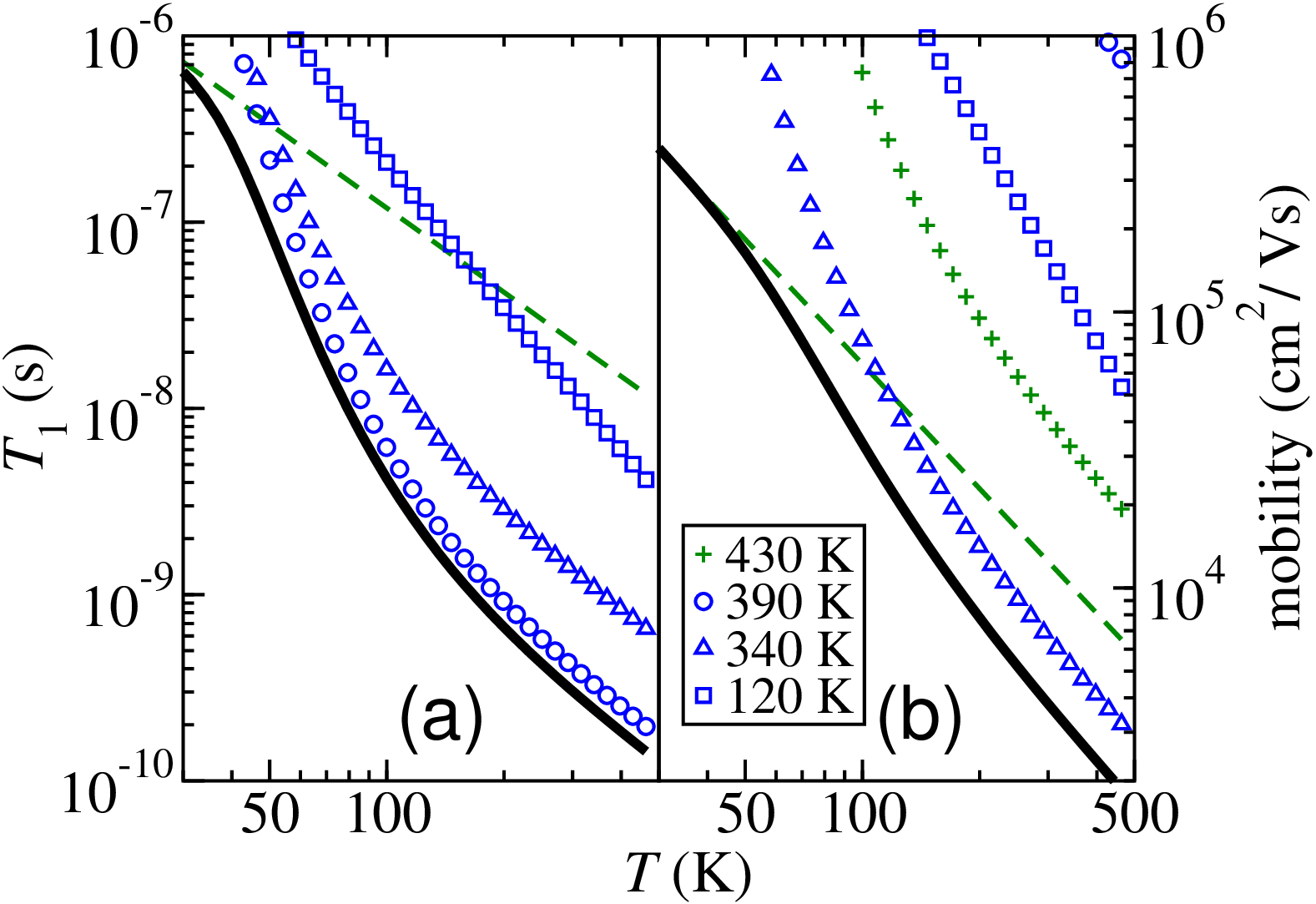}
\caption{(Color online) Electron spin-relaxation time (a) and mobility (b) in bulk germanium shown by black solid lines. The dashed line shows the intravalley acoustic phonon contribution and the plus symbol shows the intravalley optical phonon contribution. The open symbols show the intervalley $X$ processes for three effective phonon energies.}
\label{fig:ge_spin}
\end{figure}

In the spherical band approximation, we express the matrix elements in terms of the deformation potentials up to the first order in $\delta\vq=(\vk'-\vk_f)-(\vk-\vk_i)$,\cite{ferry_first-order_1976}
\begin{eqnarray}
%\lefteqn {\left|\left\langle\psi_{\vk^\prime\sigma^\prime}\left|H^{\rm ep}_\pm\right|\psi_{\vk\sigma}\right\rangle\right|^2 } \nonumber\\
M_{\sigma'\sigma}^\pm & \approx & \frac{\hbar}{2\rho V\omega}\left( D_{0,\sigma'\sigma}^2+D_{1,\sigma'\sigma}^2|\delta \vq|^2 \right)\left(n_\vq+\textstyle\frac{1}{2}\mp\frac{1}{2}\right) ,
\end{eqnarray}
where $\rho$ is the density, $V$ is the crystal volume, $\omega$ is the phonon frequency, and $\vk_i$ and $\vk_f$ are the wave numbers of the initial and final valley minima, respectively. For Si, there are six valleys on the $\Delta$ axes, e.g., $\vk_0=(2\pi/a)(0.85,0,0)$, and for Ge, there are four valleys at the $L$ points, e.g., $(\pi/a)(1,1,1)$. A spherical averaging around the valley minimum is carried out for evaluating $D_1$. 
We have assumed only the following types of matrix elements are nonzero,
$\langle\phi_{as}(\vR)|\frac{\partial H}{\partial\vR}|\phi_{ap}(\vR)\rangle$, $\langle\phi_{a'l'}(\vR')|\frac{\partial H}{\partial\vR}|\phi_{al}(\vR)\rangle$, and $\langle\phi_{a'l'}(\vR')|\frac{\partial H}{\partial\vR'}|\phi_{al}(\vR)\rangle$, and all have the same magnitude. Since the atomic potential is not explicitly known in the TBM, we can only determine the relative strengths of the processes. The overall magnitude is later determined by fixing the mobility to be $1450$ cm$^2$/Vs in Si and $3800$ cm$^2$/Vs in Ge.\cite{semiconductor_data_book} Calculated deformation potentials for different phonon processes are summarized in Table~\ref{tab:deformation_potential}. The relative strengths of different processes are consistent with the semiempirical values in the literature.\cite{ferry_first-order_1976,canali_electron_1975,jacoboni_electron_1981}

\begin{table}
\caption{ Deformation potentials. The unit for $D_0$'s is eV/\AA{}, and for $D_1$'s is eV. For intravalley acoustic processes, $D_A=3.1$ eV, $D_{A,\flip}=0.0016$ eV for Si, and $D_A=3.8$ eV, $D_{A,\flip}=0.032$ eV for Ge. For spin-flip processes, the superscript $xy$ indicates that both the initial and final valleys are in the $x$-$y$ plane, and the superscript $z$ indicates that the initial and final valleys are separated along the $z$ direction. $T_{\omega_0}$ is the effective phonon frequency in Kelvin. The deformation potentials listed here do not include the valley degeneracy of final states. The parentheses indicate that the $D_1^2$ term is negative. }
\label{tab:deformation_potential}
\begin{tabular}{ccccccccc}
\hline\hline
& Phonon & $T_{\omega_0}$ (K) & $D_0$ & $D_1$ & $D_{0,\flip}^{xy}$ & $D_{1,\flip}^{xy}$ & $D_{0,\flip}^z$ & $D_{1,\flip}^z$ \\
\hline
\multirow{7}{*}{Si} & $\Gamma_{25}^+$ & 730 & 0 & 0.34 & 0 & 0.14 & 0 & 0.19\\ \cline{2-9}
& $\Delta_2^\prime+\Delta_5$ & 700 & 4.5 & (3.2) & 0 & 0.031 & 0 & 0.01 \\
& $\Delta_1$ & 210 & 0 & 0.04 & 0 & 0.028 & 0 & 0.04 \\
& $\Delta_5$ & 140 & 0 & 2.7 & 0 & 0.009 & 0 & 0.0015 \\ \cline{2-9}
& $\Sigma_1+\Sigma_2$ & 630 & 3.2 & (2.3) & 0.03 & 0.11 & 0.044 & 0.17 \\
& $\Sigma_1+\Sigma_3$ & 500 & 3 & (1.1) & 0.018 & 0.075 & 0.026 & 0.097 \\
&$\Sigma_3+\Sigma_4$ & 210 & 0.0083 & 2.2 & 0.0083 & 0.041 & 0.0059 & 0.058 \\ \hline
\multirow{4}{*}{Ge} & $\Gamma_{25}^+$ & 430 & 3.5 & 4.6 & 0 & 0 & 0 & 0 \\ \cline{2-9}
& $X_4$ & 390 & 0.24 & 1 & 0.24 & 0.16 & 0 & 0.17 \\
& $X_1$ & 340 & 3.8 & 3.4 & 0.08 & 0.054 &0.12 & 0.089 \\
& $X_3$ & 120 & 0 & 2.7 & 0 & 0.11 & 0 & 0.1 \\
\hline\hline
\end{tabular}
\end{table}

Various electron-phonon scattering rates are computed as follows.
For the intravalley acoustic process, a linear phonon dispersion, $\omega=c|\vq|$ is used and the scattering rate including both absorption and emission is
\begin{eqnarray}
\frac{1}{\tau_A} & = & \frac{\sqrt{2}D_A^2m^{3/2}}{\pi\rho c^2\hbar^4}k_BT\left\langle\sqrt{E}\right\rangle_T ,
\end{eqnarray}
where $m=(m_Lm_T^2)^{1/3}$ is the averaged effective mass, and $c=3/(1/c_L^2+2/c_T^2)$ is the averaged speed of sound.
We use $\rho=2329$ kg/m$^3$, $m_L=0.9163m_e$, $m_T=0.1905m_e$, $c_L=8500$ m/s, and $c_T=5900$ m/s for Si, and $\rho=5323$ kg/m$^3$, $m_L=1.59m_e$, $m_T=0.0823m_e$, $c_L=4900$ m/s, and $c_T=3500$ m/s for Ge.\cite{semiconductor_data_book}
A thermal averaging of the initial electron energy is carried out with the Boltzmann distribution, appropriate for non-degenerate systems,
\begin{eqnarray}
\langle g(E)\rangle_T & \equiv & \frac{4}{3\sqrt{\pi}T^{5/2}}\int_0^\infty g(E)E^{3/2} e^{-E/T} dE .
\end{eqnarray}
In this regime the spin lifetime is independent of the carrier density; for degenerate carrier densities the spin lifetime will be shorter than for nondegenerate carrier densities as the distribution function spreads out to a larger $k$ range for which the spin mixing upon scattering will be larger.
For the intravalley optical process and each intervalley process, we use an effective phonon frequency $\omega_0$, listed in Table~\ref{tab:deformation_potential} as temperature $T_{\omega_0}$.\cite{canali_electron_1975,jacoboni_electron_1981} The zeroth-order and the first-order scattering rates are
\begin{eqnarray}
\frac{1}{\tau_{\omega_0}^{(0)}} & = & \frac{D_0^2m^{3/2}}{\sqrt{2}\pi\rho\omega_0\hbar^3}\left[ n_{\omega_0}\left\langle p_{+}^{(0)}\right\rangle_T + (n_{\omega_0}+1)\left\langle p_{-}^{(0)}\right\rangle_T \right] , \\
\frac{1}{\tau_{\omega_0}^{(1)}} & = & \frac{\sqrt{2}D_1^2m^{5/2}}{\pi\rho\omega_0\hbar^5}\left[ n_{\omega_0}\left\langle p_{+}^{(1)}\right\rangle_T + (n_{\omega_0}+1)\left\langle p_{-}^{(1)}\right\rangle_T \right] ,
\end{eqnarray}
where
\begin{eqnarray}
p_{\pm}^{(0)} & = & \sqrt{E\pm\hbar\omega_0}\,\theta(E\pm\hbar\omega_0) , \\
p_{\pm}^{(1)} & = & (2E\pm\hbar\omega_0)\sqrt{E\pm\hbar\omega_0}\,\theta(E\pm\hbar\omega_0) .
\end{eqnarray}
The final electron spin-relaxation times including all scattering processes for Si and Ge are plotted, respectively, in Figs.~\ref{fig:si_spin} and \ref{fig:ge_spin} as a function of temperature.

\begin{table}
\caption{ Selection rules with the inclusion of time-reversal symmetry without spin. For electron representations, 
$\Delta_{1t}$ is $\Delta_1$ at $\vk_0$ transformed to a valley on a perpendicular axis to $\vk_0$, and $L_{1t}$ is $L_1$ transformed from one $L$ valley to a different valley. The phonon representation $\Delta_1(2\vk_0)$ becomes $\Delta_2'(2\vk_0-(4\pi/a)(1,0,0))$ in the reduced Brillouin zone scheme used in Table~\ref{tab:deformation_potential}.}
\label{tab:selection_rules_noSO}
\begin{tabular}{clcl}
\hline\hline
\multirow{3}{*}{Si} & $\Delta_1(\vk_0)\otimes\Delta_1(\vk_0)$ & = & $\Gamma_1^+\oplus\Gamma_{12}^+$ \\
 & $\Delta_1(\vk_0)\otimes\Delta_1(-\vk_0)$ & = & $\Delta_1(2\vk_0)$  \\
 & $\Delta_1(\vk_0)\otimes\Delta_{1t}$ & = & $\Sigma_1$ \\
\hline
\multirow{2}{*}{Ge} & $L_1^+\otimes L_1^+$ & = & $\Gamma_1^+\oplus\Gamma_{25}^+$ \\
 & $L_1^+\otimes L_{1t}^+$ & = & $X_1$ \\
\hline\hline
\end{tabular}
\end{table}

\begin{table}
\caption{ Selection rules with the inclusion of time-reversal symmetry with spin. The spin-flip part of $\Delta_1$ ($\Delta_2'$ in Table~\ref{tab:deformation_potential}) in Si and of $\Gamma_{25}^+$ in Ge are also forbidden by time reversal. }
\label{tab:selection_rules}
\begin{tabular}{clcl}
\hline\hline
\multirow{3}{*}{Si} & $\Delta_6\otimes\Delta_6$ & = & $\Gamma_1^\pm\oplus\Gamma_{12}^\pm\oplus \Gamma_{15}^-\oplus \Gamma_{25}^-$ \\
 & $\Delta_6\otimes\Delta_{6}$ & = & $\Delta_1$ \\
 & $\Delta_6\otimes\Delta_{6t}$ & = & $2\Sigma_1\oplus \Sigma_2\oplus \Sigma_3$ \\
\hline
\multirow{2}{*}{Ge} & $L_6^+\otimes L_6^+$ & = & $\Gamma_1^+\oplus \Gamma_{25}^+$ \\
 & $L_6^+\otimes L_{6t}^+$ & = & $2X_1\oplus X_2\oplus X_4$ \\
\hline\hline
\end{tabular}
\end{table}

\begin{figure}
\includegraphics[width=\columnwidth]{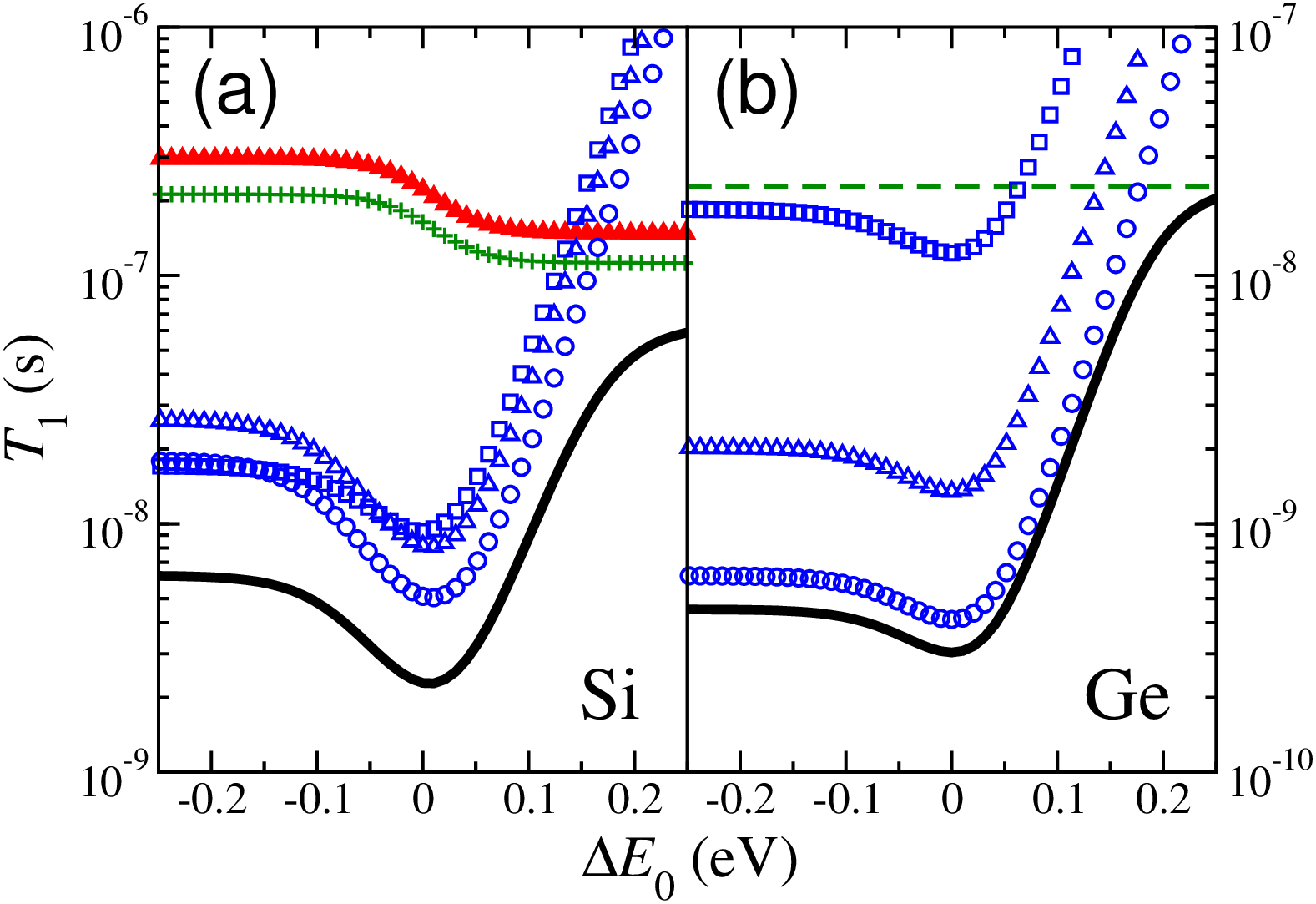}
\caption{(Color online) Spin-relaxation time at $T=300$ K as a function of valley energy shift, $\Delta E_0$, shown by black solid lines. The energy shift is the energy offset of four (three) valleys relative to the remaining two (one) in Si (Ge). The symbols in panel (a) have the same meaning as in Fig.~\ref{fig:si_spin} and in panel (b) are the same as in Fig.~\ref{fig:ge_spin}.}
\label{fig:shift}
\end{figure}

\section{Discussion}

Our results can be qualitatively understood from the selection rules, derived from symmetry, that apply to scattering processes between the valley minima, i.e., $D_0$'s.
The selection rules without spin-orbit coupling were discussed by Lax and Hopfield\cite{lax_selection_1961} and are listed in Table~\ref{tab:selection_rules_noSO}.
When electron spin is included, the irreducible representation at the conduction-band minima in Si (Ge) becomes $\Delta_6$ ($L_6^+$) instead of $\Delta_1$ ($L_1^+$).\cite{elliott_spin-orbit_1954}
We analyzed the selection rules including time-reversal symmetry using the same subgroup technique developed in Refs.~\onlinecite{lax_selection_1961,lax_influence_1962} and list the results in Table~\ref{tab:selection_rules}. The allowed phonon representations at $\vq$ (right-hand side of the equations) are obtained from the characters at $\vk$ and $\vk'$ (the two representations on the left-hand side of the equations). Details of our calculations of the selection rules are presented in the Appendix.

In Si the spin-orbit interaction mixes spin up and down by about 1\% in wave function and gives a spin-flip probability about $10^{-4}$. It can be seen from Fig.~\ref{fig:si_spin} that the $\Sigma$ phonons (also known as the $f$ processes) dominate the mobility and the spin flip near room temperature. For mobility, $\Delta$ phonons (also known as the $g$ processes) also contribute substantially, consistent with the selection rules of $\Delta_2'$ and $\Sigma_1$ in Table~\ref{tab:selection_rules_noSO}. For spin flip, the $\Sigma_2$ and $\Sigma_3$ phonons become allowed with nonzero spin-orbit interaction in addition to $\Sigma_1$ (Table~\ref{tab:selection_rules}). The intravalley optical ($\Gamma_{25}^+$) phonons and the $\Delta$ phonons remain forbidden for spin flip by time reversal and, therefore, are not as effective as the $\Sigma$ phonons. Although the coupling strength for the low-energy $\Sigma$ phonon is weaker, this is somewhat compensated by the temperature dependence of the phonon distribution and it ends up that all $\Sigma$ phonons contribute approximately the same to spin flip near room temperature.

In Ge the spin-flip probability due to spin-orbit interaction is about one order of magnitude larger, but the number of possible phonon processes is reduced. So the spin-relaxation time is only about one order of magnitude smaller than in Si as shown in Fig.~\ref{fig:ge_spin}. For mobility, the intervalley $X_1$ and the intravalley optical ($\Gamma_{25}^+$) phonons are more important due to the selection rules in Table~\ref{tab:selection_rules_noSO}. For spin flip, the high-energy ($X_4$) phonon is allowed with finite spin-orbit interaction, but the low-energy ($X_3$) phonon is still forbidden by time reversal. Although the coupling to $\Gamma_{25}^+$ phonons is as strong as the coupling to $X_1$ phonons, the spin-flip part is forbidden by time reversal and ineffective (Table~\ref{tab:selection_rules}).

Now that the structure and symmetry of the spin-relaxation mechanisms has been clarified, the analysis of the effect of strain is straightforward. Strain (or quantum confinement) breaks valley degeneracy and can eliminate multivalley scattering processes. In Si, a [001] strain can change the lowest-energy valley degeneracy from six to two (located on the same axis).  In Ge, a [111] strain can yield just a single nondegenerate valley at the conduction-band edge. We have performed a simple estimate that only takes into account the valley energy shift, $\Delta E_0$, by modifying $\pm\hbar\omega_0\rightarrow\pm\hbar\omega_0-\Delta E_0$ and the initial electron distribution in each valley. As shown in Fig.~\ref{fig:shift}, the spin lifetime averaged over all valleys can be lengthened substantially; 1\%~strain gives about $0.09$ eV shift in Si, and $0.16$ eV shift in Ge.\cite{yu_semiconductors_2010} Positive energy shift ($\Delta E_0>0$) corresponds to the configuration that four valleys out of six are shifted to higher energy in the case of Si, and three out of four in the case of Ge. Negative shift reverses the ordering. For positive shifts, the intervalley $\Sigma$ processes in Si or the $X$ processes in Ge can be completely eliminated. This tuning is eventually limited by the intervalley $\Delta$ and the intravalley optical ($\Gamma_{25}^+$) processes in Si, and by the intravalley acoustic processes in Ge. For negative shifts, the elimination of the intervalley processes is only partial, so the tuning range is much smaller.

\section{Conclusions}

We have presented a thorough symmetry analysis of the electron spin-phonon interaction processes for silicon and germanium, finding a spin lifetime for nondegenerate carriers at room temperature comparable to those in III-V semiconductors when the scattering determining the carrier mobility is dominated by phonons. However, strain or quantum confinement can lift the valley degeneracy, which lengthens the spin lifetime substantially (over an order of magnitude at room temperature).

\section*{Acknowledgements}

This work was supported in part by an ONR MURI and an ARO MURI.

\appendix

\section{Selection Rules}

In this Appendix we show the calculations of the characters of the product representations for obtaining the selection rules in Tables \ref{tab:selection_rules_noSO} and \ref{tab:selection_rules}. Each electron-phonon scattering process involves three subgroups of the wave vectors, $\vk$, $\vk'$ and $\vq$. Instead of using the intersection group formed by the elements common to all three subgroups and creating the corresponding character tables, the selection rules are derived using the existing character tables of the subgroups at these wave vectors.  The two multiplication rules, with and without time-reversal symmetry, will be presented.\cite{lax_selection_1961,lax_influence_1962} We choose to compute the characters of the phonon representations at $\vq$ from the products of the electron representations at $\vk$ and $\vk'$. Once the characters at $\vq$ are obtained, the decomposition into irreducible representations is done in the usual way using the subgroup of the wave vector $\vq$.

\begin{table}[ht]
\caption{Double group characters at $\Delta = k_0(1,0,0)$.}
\label{tab:Dtable}
\begin{tabular}{c|ccccc|cc}
\hline\hline
Class & $\Delta_{1}$ & $\Delta_{2}$ & $\Delta_2'$ & $\Delta_1'$ & $\Delta_{5}$ & $\Delta_6$ & $\Delta_7$ \\
\hline
$(E|0)$          & 1 &  1   & 1 &  1   &  2   &  2   &  2   \\
$(\bar E|0)$     & 1 &  1   & 1 &  1   &  2   & $-2$ & $-2$ \\
$(C_4^2|0),(\bar C_4^2|0)$  & 1 &  1   & 1 &  1   & $-2$ & 0 & 0 \\
$2(C_4|\tau)$      & $\lambda$ & $-\lambda$ & $-\lambda$ & $\lambda$ & 0 & $\sqrt{2}\lambda$ & $-\sqrt{2}\lambda$ \\
$2(\bar C_4|\tau)$ & $\lambda$ & $-\lambda$ & $-\lambda$ & $\lambda$ & 0 & $-\sqrt{2}\lambda$ & $\sqrt{2}\lambda$ \\
$2(iC_4^2|\tau),2(i\bar C_4^2|\tau)$ & $\lambda$ & $\lambda$ & $-\lambda$ & $-\lambda$ & 0 & 0 & 0 \\
$2(iC_2|0),2(i\bar C_2|0)$      & 1 & $-1$ & 1 & $-1$ &  0  & 0 & 0 \\
\hline\hline
\end{tabular}\\
$\lambda = e^{i k_0 a/4}$, $\bar E$ is $2\pi$ rotation, $N_{\Delta\text{ star }\Delta}(C)=1$, \\
$\Delta_6 = \Delta_1 \otimes D_{1/2}$, $\Delta_7 = \Delta_2 \otimes D_{1/2}$
\end{table}

\begin{table}[ht]
\caption{Double group characters at $L = (\pi/a)(1,1,1)$.}
\label{tab:Ltable}
\begin{tabular}{c|ccc|ccc}
\hline\hline
Class & $L_1^\pm$ & $L_2^\pm$ & $L_3^\pm$ & $L_4^\pm$ & $L_5^\pm$ & $L_6^\pm$ \\
\hline
$(E|0)$            &  1  &  1   &  2   &  1   &  1   &  2   \\
$(\bar E|0)$       &  1  &  1   &  2   & $-1$ & $-1$ & $-2$ \\
$3(C_2|\tau)$      &  1  & $-1$ &  0   &  $i$ & $-i$ &  0   \\
$3(\bar C_2|\tau)$ &  1  & $-1$ &  0   & $-i$ &  $i$ &  0   \\
$2(C_3|0)$         &  1  &  1   & $-1$ & $-1$ & $-1$ &  1   \\
$2(\bar C_3|0)$    &  1  &  1   & $-1$ &  1   &  1   & $-1$ \\
$(i|\tau)Z$        &  & $\pm\chi(Z)$ & & & $\pm\chi(Z)$ &  \\
%$(i|\tau)$         & $\pm\eta$ & $\pm\eta$ & $\pm 2\eta$ & $\pm\eta$ & $\pm\eta$ & $\pm 2\eta$ \\
%$(\bar i|\tau)$    & $\pm\eta$ & $\pm\eta$ & $\pm 2\eta$ & $\mp\eta$ & $\mp\eta$ & $\mp 2\eta$ \\
%$2(iC_3|\tau)$     & $\pm\eta$ & $\pm\eta$ & $\mp\eta$   & $\mp\eta$ & $\mp\eta$ & $\pm\eta$ \\
%$2(i\bar C_3|\tau)$& $\pm\eta$ & $\pm\eta$ & $\mp\eta$   & $\pm\eta$ & $\pm\eta$ & $\mp\eta$ \\
%$3(iC_2|0)$        & $\pm 1$   & $\mp 1$   &  0   &  $i$ & $-i$ &  0   \\
%$3(i\bar C_2|0)$   & $\pm 1$   & $\mp 1$   &  0   & $-i$ &  $i$ &  0   \\
\hline\hline
\end{tabular} \\
$Z$ can be any of the six classes shown above.\\
$L_6^\pm = L_1^\pm \otimes D_{1/2}= L_2^\pm \otimes D_{1/2}$
\end{table}

We first carry out calculations without time-reversal symmetry. To obtain the product character at $\vq$ for each class $C$ using the characters at $\vk$ and at $\vk'$, the usual character multiplication rules are modified as follows.
\begin{eqnarray}
\chi^{i\otimes j}_\vq(C) & = & \chi^i_{\vk'}(C)\left[\chi^{j}_{\vk}(C)\right]^*N_{\vq \text{ star }\vk}(C) \;,
\end{eqnarray}
where $i$ and $j$ are the irreducible representations, and $N_{\vq \text{ star }\vk}(C)$ is the number of wave vectors that are unchanged by the class $C$, out of a set of nonequivalent $\vk$ points. This set of nonequivalent $\vk$ points is generated by the subgroup of $\vq$ and is called ``$\vq$ star of $\vk$.'' The product character is simply zero if $C$ is not a common class of all three subgroups.

\begin{table}[ht]
\caption{Group characters at $\Gamma = (0,0,0)$.}
\label{tab:Gtable}
\begin{tabular}{c|ccccc|c|c}
\hline
Class & $\Gamma_{1}^\pm$ & $\Gamma_{2}^\pm$ & $\Gamma_{12}^\pm$ & $\Gamma_{15}^\pm$ & $\Gamma_{25}^\pm$ & $N_{\Gamma\text{ star }\Delta}$ & $N_{\Gamma\text{ star }L}$ \\
\hline\hline
$(E|0)$       & 1 &  1   &  2   &  3   &  3   & 6 & 4 \\
$3(C_4^2|0)$  & 1 &  1   &  2   & $-1$ & $-1$ & 2 & 0 \\
$6(C_4|\tau)$ & 1 & $-1$ &  0   &  1   & $-1$ & 2 & 0 \\
$6(C_2|\tau)$ & 1 & $-1$ &  0   & $-1$ &  1   & 0 & 2 \\
$8(C_3|0)$    & 1 &  1   & $-1$ &  0   &  0   & 0 & 1 \\

$(i|\tau)$       & $\pm 1$ & $\pm 1$ & $\pm 2$ & $\pm 3$ & $\pm 3$ & 0 & 4 \\
$3(iC_4^2|\tau)$ & $\pm 1$ & $\pm 1$ & $\pm 2$ & $\mp 1$ & $\mp 1$ & 4 & 0 \\
$6(iC_4|0)$      & $\pm 1$ & $\mp 1$ & 0       & $\pm 1$ & $\mp 1$ & 0 & 0 \\
$6(iC_2|0)$      & $\pm 1$ & $\mp 1$ & 0       & $\mp 1$ & $\pm 1$ & 2 & 2 \\
$8(iC_3|\tau)$   & $\pm 1$ & $\pm 1$ & $\mp 1$ & 0       & 0       & 0 & 1 \\
\hline\hline
\end{tabular}
\end{table}

\begin{table}[ht]
\caption{Group characters at $\Sigma = k_0(1,1,0)$.}
\label{tab:Stable}
\begin{tabular}{c|cccc|c}
\hline\hline
Class & $\Sigma_1$ & $\Sigma_2$ & $\Sigma_3$ & $\Sigma_4$ & $N_{\Sigma\text{ star }\Delta}$ \\
\hline
$(E|0)$    & 1 &  1   & 1 &  1   & 2 \\
$(C_2|\tau)$    & $\lambda^2$ &  $\lambda^2$ & $-\lambda^2$ & $-\lambda^2$ & 0 \\
$(iC_4^2|\tau)$ & $\lambda^2$ & $-\lambda^2$ & $-\lambda^2$ &  $\lambda^2$ & 2 \\
$(iC_2|0)$ & 1 & $-1$ & 1 & $-1$ & 0 \\
\hline\hline
\end{tabular}\\
$\lambda^2 = e^{i k_0 a/2}$
\end{table}

\begin{table}[ht]
\caption{Group characters at $X = (2\pi/a)(1,0,0)$.}
\label{tab:Xtable}
\begin{tabular}{c|cccc|c}
\hline\hline
Class & $X_1$ & $X_2$ & $X_3$ & $X_4$ & $N_{X\text{ star }L}$ \\
\hline
$(E|0)$                         &  2   &  2   &  2   &  2   & 4 \\
$(C_4^2|0)$                     &  2   &  2   & $-2$ & $-2$ & 0 \\
$(C_2|\tau),(C_2'|\tau+t_{xy})$ &  0   &  0   & $-2$ &  2   & 2 \\
$2(iC_2|0)$                     &  2   & $-2$ &  0   &  0   & 2 \\
%$(E|t_{xy})Z$ &  & $-\chi(Z)$ & & & $N(Z)$ \\
$(E|t_{xy})$                    & $-2$ & $-2$ & $-2$ & $-2$ & 4 \\
$(C_4^2|t_{xy})$                & $-2$ & $-2$ &  2   &  2   & 0 \\
$(C_2|\tau+t_{xy}),(C_2'|\tau)$ &  0   &  0   &  2   & $-2$ & 2 \\
$2(iC_2|t_{xy})$                & $-2$ &  2   &  0   &  0   & 2 \\
\hline\hline
\end{tabular}\\
%$Z$ can be any of the four classes shown above.
$t_{xy}=(a/2)(1,1,0)$
\end{table}

Listed here are the character tables for the double groups representing the symmetry of the electron states at $\Delta$ for Si (Table~\ref{tab:Dtable}) and at $L$ for Ge (Table~\ref{tab:Ltable}), and the tables for the groups at $\Delta$, $\Gamma$, $\Sigma$, and $X$ for phonons (Tables~\ref{tab:Dtable} and \ref{tab:Gtable}--\ref{tab:Xtable}). The numbers of invariant $\vq$-star-$\vk$ points, $N_{\vq\text{ star }\vk}(C)$, are listed in the tables for phonons. Note that the group at $X$ has 14 irreducible representations, 14 classes, and 32 elements. Only the 4 physically admissible representations and 8 classes are listed in the table. The other 6 classes are not relevant because they have zero character for the 4 physical representations. Further simplification is possible because one can work with just 4 classes; the other 4 classes, corresponding to an additional lattice translation, $t_{xy}=(a/2)(1,1,0)$, have exactly the opposite characters. 

The relevant characters of the product representations are listed in Tables~\ref{tab:prodDDG}--\ref{tab:prodDDS} for Si and in Tables~\ref{tab:prodLLG}--\ref{tab:prodLLX} for Ge. Because the space group for the diamond structure is non-symmorphic, some of the group elements contain a sublattice translation, $\tau=(a/4)(1,1,1)$. Note that the complex phase factors ($\lambda$) due to this translation in some characters should always cancel out in the final character product ($\vk'=\vk+\vq$) and not affect the selection rules. We will ignore these explicit phase factors in our character product tables. However, care needs to be taken with the product classes of $L$ and $L_t$ when comparing to the classes at $X$. The product of two group elements with $\tau$ can result in a lattice translation, $t_{xy}$, which is explicitly included in the subgroup of $X$, but not in the subgroup of $L$. All physical representations at $X$ are odd under $t_{xy}$. The characters at $L=(\pi/a)(1,1,1)$ are odd under $t_{xy}$, but are even at $L_t=(\pi/a)(-1,1,1)$. Therefore, the product class for $(C_2|\tau)$ at $L$ and at $L_{t}$ is actually $(C_2|\tau+t_{xy})$ at $X$. This is why the product of $L_1$ and $L_{1t}$ contains $X_3$, but not $X_4$.\cite{lax_selection_1961}

Time reversal can add additional constraints to the selection rules if there exists a group element that connects the time-reversed process ($-\vk'\rightarrow -\vk$) to the original process ($\vk\rightarrow\vk'$). That is, we need an element $Q$ from the subgroup of $\vq$ that interchanges $\vk$ and $-\vk'$.
To incorporate the time reversal symmetry, the character multiplication rule is modified to the symmetric or antisymmetric combination of the characters of the original process and the process connected via $Q$,
\begin{eqnarray}
\chi_{\vq\pm}^{i\otimes j}(C) & = & \frac{1}{2}\left[\chi_\vq^{i\otimes j}(C)\pm\chi_\vk^j(C^2)N_{\vq\text{ star }\vk}(QC)\right] \;,
\end{eqnarray}
where $C^2$ is the class of the square of elements in $C$ and $Q$ is the element that interchanges $\vk$ and $-\vk'$. The positive (symmetric) sign is used in the cases without spin and the negative (antisymmetric) sign is used with spin. Again, all the relevant characters for the product representations are listed in Tables~\ref{tab:prodDDG}--\ref{tab:prodLLX}. The element $Q$ can be identified as the first entry in the column of $QC$ when $C$ is the identity class. For $\Delta$ phonons in Si and $\Gamma$ phonons in Ge, $Q$ is simply the identity element.
Upon closer examination, we also found that the spin-flip part of the $\Delta_1$ phonon process in Si and of the $\Gamma_{25}^+$ phonon process in Ge are forbidden because the final state is exactly the time reverse of the initial state (e.g., $|\alpha\rangle=|\vk\uparrow\rangle$ and $|\hat T\alpha\rangle = |-\vk\downarrow\rangle$),
\begin{eqnarray}
\langle\hat T\alpha |\hat H_{\rm ep}|\alpha\rangle & = & \langle\hat T\alpha |\hat T \hat H_{\rm ep}^\dagger\hat T^{-1}|\hat T^2\alpha\rangle \;,
\end{eqnarray}
where $\hat T$ is the time-reversal operator, $\hat T^2=-\hat 1$ in the presence of half-integer spin, and $\hat H_{\rm ep}=\hat T \hat H_{\rm ep}^\dagger\hat T^{-1}$ is the time-reversal invariant electron-phonon interaction Hamiltonian.

%si_spin-v2.bbl
%Merlin.mbs v4.21 2009-07-09.

\begin{table*}[ht]
\caption{Characters of the product representations at $\Gamma$.}
\label{tab:prodDDG}
\begin{tabular}{c|cc|cccc|cccc}
\hline\hline
$C$ & $QC$ & $N(QC)$ & $C^2$ & $\chi_{\vk_0}^{\Delta_1}(C^2)$ & $\chi_{\Gamma}^{\Delta_1\otimes\Delta_1}$ & $\chi_{\Gamma+}^{\Delta_1\otimes\Delta_1}$ & $C^2$ & $\chi_{\vk_0}^{\Delta_6}(C^2)$ & $\chi_{\Gamma}^{\Delta_6\otimes\Delta_6}$ & $\chi_{\Gamma-}^{\Delta_6\otimes\Delta_6}$ \\ \hline
$(E|0)$         & $(i|\tau)$      & 0 & $(E|0)$     & 1 & 6 & 3 & $(E|0)$                    &  2   & 24 & 12 \\
$(C_4^2|0)$     & $(iC_4^2|\tau)$ & 4 & $(E|0)$     & 1 & 2 & 3 & $(\bar E|0)$               & $-2$ &  0 & 4 \\
$(C_4|\tau)$    & $(iC_4|0)$      & 0 & $(C_4^2|0)$ & 1 & 2 & 1 & $(C_4^2|0),(\bar C_4^2|0)$ &  0   &  4 & 2 \\
$(C_2|\tau)$    & $(iC_2|0)$      & 2 & $(E|0)$     & 1 & 0 & 1 & $(\bar E|0)$               & $-2$ &  0 & 2 \\
$(i|\tau)$      & $(E|0)$         & 6 & $(E|0)$     & 1 & 0 & 3 & $(E|0)$                    &  2   &  0 & $-6$ \\
$(iC_4^2|\tau)$ & $(C_4^2|0)$     & 2 & $(E|0)$     & 1 & 4 & 3 & $(\bar E|0)$               & $-2$ &  0 & 2 \\
$(iC_4|0)$      & $(C_4|\tau)$    & 2 & $(C_4^2|0)$ & 1 & 0 & 1 & $(C_4^2|0),(\bar C_4^2|0)$ &  0   &  0 & 0 \\
$(iC_2|0)$      & $(C_2|\tau)$    & 0 & $(E|0)$     & 1 & 2 & 1 & $(\bar E|0)$               & $-2$ &  0 & 0 \\
\hline\hline
\end{tabular}\\
Only classes that have non-trivial characters are shown.\\
$\Delta_1\otimes\Delta_1$ = $\Gamma_1^+\oplus\Gamma_{12}^+\oplus\Gamma_{15}^-$ and
$\Delta_6\otimes\Delta_6 = \Gamma_1^\pm\oplus\Gamma_{12}^\pm\oplus 2\Gamma_{15}^\pm\oplus\Gamma_{25}^\pm$ without time-reversal symmetry
\end{table*}

\begin{table*}[ht]
\caption{Characters of the product representations at $\Delta=2\vk_0$.}
\label{tab:prodDDD}
\begin{tabular}{c|cccc|cccc}
\hline\hline
$C$ & $C^2$ & $\chi_{\vk_0}^{\Delta_1}(C^2)$ & $\chi_{\Delta}^{\Delta_1\otimes\Delta_1}$ & $\chi_{\Delta+}^{\Delta_1\otimes\Delta_1}$ & $C^2$ & $\chi_{\vk_0}^{\Delta_6}(C^2)$ & $\chi_{\Delta}^{\Delta_6\otimes\Delta_6}$ & $\chi_{\Delta-}^{\Delta_6\otimes\Delta_6}$ \\
\hline
$(E|0)$         & $(E|0)$ & 1 & 1 & 1 & $(E|0)$                    &  2   & 4 & 1 \\
$(C_4^2|0)$     & $(E|0)$ & 1 & 1 & 1 & $(\bar E|0)$               & $-2$ & 0 & 1 \\
$(C_4|\tau)$    & $(C_4^2|0)$ & 1 & 1 & 1 & $(C_4^2|0),(\bar C_4^2|0)$ &  0   & 2 & 1 \\
$(iC_4^2|\tau)$ & $(E|0)$ & 1 & 1 & 1 & $(\bar E|0)$               & $-2$ & 0 & 1 \\
$(iC_2|0)$      & $(E|0)$ & 1 & 1 & 1 & $(\bar E|0)$               & $-2$ & 0 & 1 \\
\hline\hline
\end{tabular}\\
$Q$ is the identity element. $\Delta_6\otimes\Delta_6 = \Delta_1\oplus\Delta_1^\prime\oplus\Delta_5$ without time-reversal symmetry
\end{table*}

\begin{table*}[ht]
\caption{Characters of the product representations at $\Sigma=k_0(1,1,0)$.}
\label{tab:prodDDS}
\begin{tabular}{c|cc|cccc|cccc}
\hline\hline
$C$ & $QC$ & $N(QC)$ & $C^2$ & $\chi_{\vk_0}^{\Delta_1}(C^2)$ & $\chi_{\Sigma}^{\Delta_1\otimes\Delta_{1t}}$ & $\chi_{\Sigma+}^{\Delta_1\otimes\Delta_{1t}}$ & $C^2$ & $\chi_{\vk_0}^{\Delta_6}(C^2)$ & $\chi_{\Sigma}^{\Delta_6\otimes\Delta_{6t}}$ & $\chi_{\Sigma-}^{\Delta_6\otimes\Delta_{6t}}$ \\
\hline
$(E|0)$         & $(C_2|\tau)$    & 0 & $(E|0)$ & 1 & 2 & 1 & $(E|0)$      &  2   & 8 & 4 \\
$(C_2|\tau)$    & $(E|0)$         & 2 & $(E|0)$ & 1 & 0 & 1 & $(\bar E|0)$ & $-2$ & 0 & 2 \\
$(iC_4^2|\tau)$ & $(iC_2|0)$      & 0 & $(E|0)$ & 1 & 2 & 1 & $(\bar E|0)$ & $-2$ & 0 & 0 \\
$(iC_2|0)$      & $(iC_4^2|\tau)$ & 2 & $(E|0)$ & 1 & 0 & 1 & $(\bar E|0)$ & $-2$ & 0 & 2 \\
\hline\hline
\end{tabular}\\
$\Delta_1\otimes\Delta_{1t} = \Sigma_1\oplus\Sigma_4$ and $\Delta_6\otimes\Delta_{6t} = 2\Sigma_1\oplus 2\Sigma_2\oplus 2\Sigma_3\oplus 2\Sigma_4$ without time-reversal symmetry
\end{table*}

\begin{table*}[ht]
\caption{Characters of the product representations at $\Gamma$.}
\label{tab:prodLLG}
\begin{tabular}{c|cccc|cccc}
\hline\hline
$C$ & $C^2$ & $\chi_{L}^{L_1^+}(C^2)$ & $\chi_{\Gamma}^{L_1^+\otimes L_1^+}$ & $\chi_{\Gamma+}^{L_1^+\otimes L_1^+}$ & $C^2$ & $\chi_{L}^{L_6^+}(C^2)$ & $\chi_{\Gamma}^{L_6^+\otimes L_6^+}$ & $\chi_{\Gamma-}^{L_6^+\otimes L_6^+}$ \\
\hline
$(E|0)$       & $(E|0)$   & 1 & 4 & 4 & $(E|0)$        &  2   & 16 & 4 \\
$(C_2|\tau)$  & $(E|0)$   & 1 & 2 & 2 & $(\bar E|0)$   & $-2$ &  0 & 2 \\
$(C_3|0)$     & $(C_3|0)$ & 1 & 1 & 1 & $(\bar C_3|0)$ & $-1$ &  1 & 1 \\
$(i|\tau)$    & $(E|0)$   & 1 & 4 & 4 & $(E|0)$        &  2   & 16 & 4 \\
$(iC_2|0)$    & $(E|0)$   & 1 & 2 & 2 & $(\bar E|0)$   & $-2$ &  0 & 2 \\
$(iC_3|\tau)$ & $(C_3|0)$ & 1 & 1 & 1 & $(\bar C_3|0)$ & $-1$ &  1 & 1 \\
\hline\hline
\end{tabular}\\
$Q$ is the identity element. Only classes that have non-trivial characters are shown.\\
$L_6^+\otimes L_6^+ = \Gamma_1^+\oplus\Gamma_2^+\oplus\Gamma_{12}^+\oplus 2\Gamma_{15}^+\oplus 2\Gamma_{25}^+$ without time-reversal symmetry
\end{table*}

\begin{table*}[ht]
\caption{Characters of the product representations at $X$.}
\label{tab:prodLLX}
\begin{tabular}{c|cc|cccc|cccc}
\hline\hline
$C$ & $QC$ & $N(QC)$ & $C^2$ & $\chi_{L}^{L_1^+}(C^2)$ & $\chi_{X}^{L_1^+\otimes L_{1t}^+}$ & $\chi_{X+}^{L_1^+\otimes L_{1t}^+}$ & $C^2$ & $\chi_{L}^{L_6^+}(C^2)$ & $\chi_{X}^{L_6^+\otimes L_{6t}^+}$ & $\chi_{X-}^{L_6^+\otimes L_{6t}^+}$ \\
\hline
$(E|0)$                         & $(C_4^2|0)$                     & 0 & $(E|0)$ & 1 &  4   &  2   & $(E|0)$      &   2    & 16 & 8 \\
$(C_4^2|0)$                     & $(E|0)$                         & 4 & $(E|0)$ & 1 &  0   &  2   & $(\bar E|0)$ & $-2$ &  0 & 4 \\
$(C_2|\tau),(C_2'|\tau+t_{xy})$ & $(C_2'|\tau),(C_2|\tau+t_{xy})$ & 2 & $(E|0)$ & 1 & $-2$ &  0   & $(\bar E|0)$ & $-2$ &  0 & 2 \\
$(iC_2|0)$                      & $(iC_2|0)$                      & 2 & $(E|0)$ & 1 &  2   &  2   & $(\bar E|0)$ & $-2$ &  0 & 2 \\
%$(E|t_{xy})$                    & $(C_4^2|t_{xy})$                & 0 & $(E|0)$ & $-1$ & $-4$ & $-2$ & $(E|0)$      & $-2$    & $-16$ & -8 \\
%$(C_4^2|t_{xy})$                & $(E|t_{xy})$                    & 4 & $(E|0)$ & $-1$ &  0   & $-2$ & $(\bar E|0)$ & $2$ &  0 & -4 \\
%$(C_2|\tau+t_{xy}),(C_2'|\tau)$ & $(C_2'|\tau+t_{xy}),(C_2|\tau)$ & 2 & $(E|0)$ & $-1$ &  2   &  0   & $(\bar E|0)$ & $2$ &  0 & -2 \\
%$(iC_2|t_{xy})$                 & $(iC_2|t_{xy})$                 & 2 & $(E|0)$ & $-1$ & $-2$ & $-2$ & $(\bar E|0)$ & $2$ &  0 & -2 \\
\hline\hline
\end{tabular}\\
Four other non-trivial classes that have exactly the opposite characters are not shown. \\
$L_1\otimes L_{1t} = X_1\oplus X_3$ and $L_6^+\otimes L_{6t}^+ = 2X_1\oplus 2X_2\oplus 2X_3\oplus 2X_4$ without time-reversal symmetry
\end{table*}

\end{document}